# A New Digital Watermarking Algorithm Using Combination of Least Significant Bit (LSB) and Inverse Bit

Abdullah Bamatraf, Rosziati Ibrahim and Mohd. Najib Mohd. Salleh

**Abstract**—In this paper, we introduce a new digital watermarking algorithm using least significant bit (LSB). LSB is used because of its little effect on the image. This new algorithm is using LSB by inversing the binary values of the watermark text and shifting the watermark according to the odd or even number of pixel coordinates of image before embedding the watermark. The proposed algorithm is flexible depending on the length of the watermark text. If the length of the watermark text is more than ((MxN)/8)-2 the proposed algorithm will also embed the extra of the watermark text in the second LSB. We compare our proposed algorithm with the 1-LSB algorithm and Lee's algorithm using Peak signal-to-noise ratio (PSNR). This new algorithm improved its quality of the watermarked image. We also attack the watermarked image by using cropping and adding noise and we got good results as well.

◆ ————————————

## 1  INTRODUCTION

Privacy is the ability of an individual or group to insulate them or information about themselves and thereby reveal them selectively [1]. Data privacy is the relationship between collection and dissemination of data, technology, the public anticipation of privacy, and the legal issues [2]. Data privacy or data protection has become increasingly important as more and more systems are connected to the internet [2]. Watermarking is a pattern of bits inserted into a digital image, audio or video file that specifies the file's copyright information such author, rights and so on [3]. Thus, watermarking approach is used to make sure of the protection of the data. However, watermarking is also designed to be completely invisible. The actual bits representing the watermark must be scattered throughout the file in such a way that they cannot be identified and tampered [4]. Thus, the watermarking must be robust enough so that it can withstand normal changes to the file such as attacking by adding noise [5].

Contrast to printed watermarks, digital watermarking is a technique where bits of information are embedded in such a way that is completely invisible [6]. The problem with the traditional way of printing logos or names is that they may be easily tampered or duplicated. In digital watermarking, the actual bits are scattered in the image in such a way that they cannot be identified and show resilience against attempts to remove the hidden data [7].

Media watermarking research is a very active area and digital image watermarking became an interesting protection measure and got the attention of many researchers since the early 1990s [8].

The rest of this paper is organized as follows: Section 2 describes the related work and LSB. Section 3 discusses the proposed algorithm. Results and discussion is given in Section 4. The PSNR and its results are shown in section 5. Discussing attacks on the watermarked images in section 6 and finally, conclusion will be presented in Section 7.

## 2  RELATED WORK

In this section we will look into the review of digital watermarks used for images. It describes the previous work which had been done on digital watermarking by using LSB technique and other techniques, including the analysis of various watermarking schemes and their results.

Gaurav Bhatnagar et al [9], presented a semi-blind reference watermarking scheme based on discrete wavelet transform (DWT) and singular value decomposition (SVD) for copyright protection and authenticity. Their watermark was a gray scale logo image. For watermark embedding, their algorithm transformed the original image into wavelet domain and a reference sub-image is formed using directive contrast and wavelet coefficients. Then, their algorithm embedded the watermark into

————————————————

- A. Bamatraf is with the universiti Tun Hussein Onn Malaysia, 86400 Batu Pahat, Johor, Malaysia. E-mail: abdom45@hotmail.com.
- R. Ibrahim is with the universiti Tun Hussein Onn Malaysia, 86400 Batu Pahat, Johor, Malaysia. E-mail: rosziati@uthm.edu.my.
- M. N. M. Salleh is with the universiti Tun Hussein Onn Malaysia, 86400 Batu Pahat, Johor, Malaysia. E-mail: najib@uthm.edu.my.

reference image by modifying the singular values of reference image using the singular values of the watermark.

Hao Luo et al [10], proposed a self-embedding watermarking scheme for digital images. In their proposed algorithm they used the cover image as a watermark. It generates the watermark by halftoning the host image into a halftone image. Then, the watermark is permuted and embedded in the LSB of the host image. The watermark is retrieved from the LSB of the suspicious image and inverse permuted.

Wen-Chao Yang et al [11] used the PKI (Public-Key Infrastructure), Public-Key Cryptography and watermark techniques to design a novel testing and verifying method of digital images. The main idea of their paper is to embed encryption watermarks in the least significant bit (LSB) of cover images.

Hong Jie He et al [12], proposed a wavelet-based fragile watermarking scheme for secure image authentication. In their proposed scheme, they generated the embedded watermark using the discrete wavelet transform (DWT), and then they elaborated security watermark by scrambling encryption is embedded into the least significant bit (LSB) of the host image.

Sung-Cheal Byun et al [13], proposed a fragile watermarking scheme for authentication of images. They used singular values of singular value decomposition (SVD) of images to check the integrity of images. In order to make authentication data, the singular values are changed to the binary bits using modular arithmetic. Then, they inserted the binary bits into the least significant bits (LSBs) of the original image. The pixels to be changed are randomly selected in the original image.

Gil-Je Lee et al [14] presented a new LSB digital watermarking scheme by using random mapping function. The idea of their proposed algorithm is embedding watermark randomly in the coordinates of the image by using random function to be more robust than the traditional LSB technique.

Saeid Fazli et al [15] presented trade-off between imperceptibility and robustness of LSB watermarking using SSIM Quality Metrics. In their algorithm, they put significant bit-planes of the watermark image instead of lower bit-planes of the asset picture.

Debjyoti Basu et al [16] proposed Bit Plane Index Modulation (BPIM) based fragile watermarking scheme for authenticating RGB color image. By embedding R, G, B component of watermarking image in the R, G, B component of original image, embedding distortion is minimized by adopting least significant bit (LSB) alteration scheme. Their proposed method consists of encoding and decoding methods that can provide public detection capabilities in the absences of original host image and watermark image.

To overcome the drawback of existing techniques, we would like to introduce a new alternative technique by inserting watermark text in grayscale images by using watermarking approach.

## 3 REVIEW OF LSB

In a digital image, information can be inserted directly into every bit of image information or the more busy areas of an image can be calculated so as to hide such messages in less perceptible parts of an image [14],[17].

Tirkel et al [18] were one of the first used techniques for image watermarking. Two techniques were presented to hide data in the spatial domain of images by them. These methods were based on the pixel value's Least Significant Bit (LSB) modifications. The algorithm proposed by Kurah and McHughes [19] to embed in the LSB and it was known as image downgrading [20].

An example of the less predictable or less perceptible is Least Significant Bit insertion. This section explains how this works for an 8-bit grayscale image and the possible effects of altering such an image. The principle of embedding is fairly simple and effective. If we use a grayscale bitmap image, which is 8- bit, we would need to read in the file and then add data to the least significant bits of each pixel, in every 8-bit pixel.

In a grayscale image each pixel is represented by 1 byte consist of 8 bits. It can represent 256 gray colors between the black which is 0 to the white which is 255. The principle of encoding uses the Least Significant Bit of each of these bytes, the bit on the far right side.

If data is encoded to only the last two significant bits (which are the first and second LSB) of each color component it is most likely not going to be detectable; the human retina becomes the limiting factor in viewing pictures [17].

For the sake of this example only the least significant bit of each pixel will be used for embedding information. If the pixel value is 138 which is the value 10000110 in binary and the watermark bit is 1, the value of the pixel will be 10000111 in binary which is 139 in decimal. In this example we change the underline pixel.

## 4 PROPOSED METHOD

Based on LSB technique, we propose a new watermarking algorithm. Most of researchers have proposed the first LSB but our proposed watermarking algorithm is inversing the watermark text and embedding it in different order in the traditional LSB. This is because of the security reason. So, no one will expect that the hidden watermark text in different order and it is inversed. Figure 1 shows the framework of the proposed method. First, we select the image which is a grayscale image and we will transfer the watermark to binary value after typing it. Then, we embed the watermark in the image using the proposed algorithm. Figure 2 shows the embedding algorithm. Then, we will get the watermarked image. Then, the receiver will retrieve the watermark back. The watermark will be extracted from the watermarked image. Figure 3 shows the extracting algorithm.

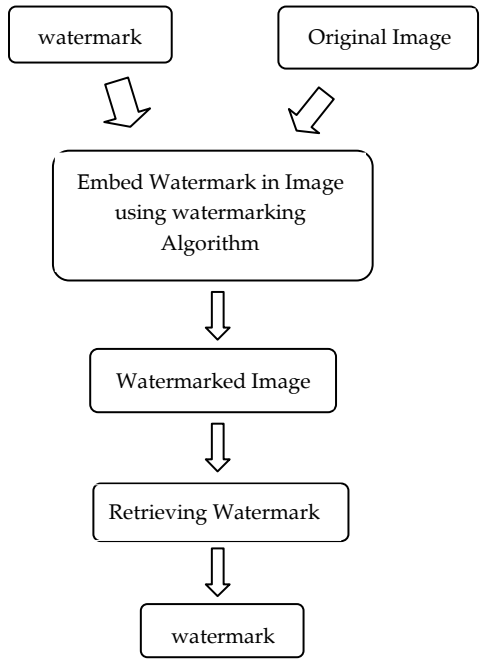

Fig. 1. The framework of the proposed method

## 4.1 Embedding Algorithm

In this section, we describe the embedding algorithm. After we select the image and type the watermark text, we transfer the watermark text to binary values and determine the coordinates of the image which the watermark will be embedded in. First, we will embed the length of the watermark text in sixteen pixels starting from the first coordinate which we select until we embed it in the sixteen pixels in the 1st LSB. Based on the length of the watermark text, we can know how many copies it will be embedded and if we are going to embed in the 2nd LSB. Before the watermark will be embedded in the image in the 1st LSB, it will be inversed and we will change the order of embedding. So, if the coordinate X is even, it will subtract 1 from X and if X is odd, it will add 1 to X. Then, watermarked image will be produced and it will be saved. Figure 2 shows the embedding algorithm.

## 4.2 Extracting Algorithm

In this section, we will describe the extracting algorithm. After receiving the watermarked image, we will get the length of the watermark text from the 1st LSB in the sixteen pixels starting from the determined coordinates until we get it from the sixteen pixels. After getting the length, we can know how many copies the sender has embedded. So, we can choose any copy to be displayed and also we can get the embedded watermark in the second LSB, if there. Then, the algorithm will check if the coordinate X is even it will subtract 1 from X of if the coordinate X is odd, it will add 1 to X. After that, it will get the watermark bit which will be inversed. Finally, the watermark bits will be transferred to characters which will be shown as the watermark text. Figure 3 shows the extracting algorithm.

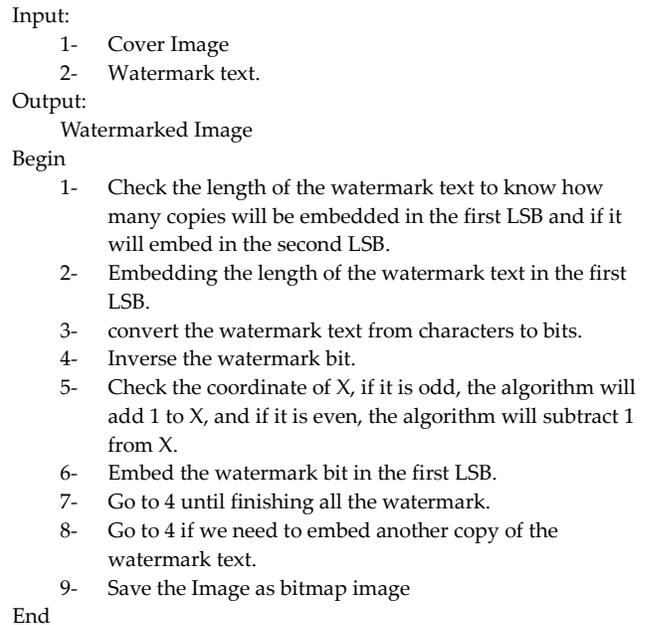

Fig. 2. Embedding Algorithm

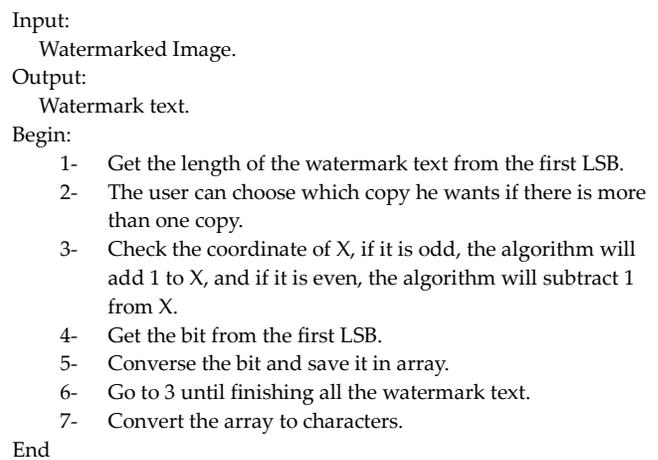

Fig. 3. The extracting algorithm

## 5   EXPERIMENTAL RESULTS AND DISCUSSION

In our experimental results, we have four 512x512 grayscale BMP images are shown in figure 4 were used as cover images. The size of every cover image is 257 kilo-bytes. We have tested different watermark text in different places of the pixels. As it is known that every pixel in the grayscale image contains 8 bits. And every bit have value, for example the first bit in the right has the value 1 and it is called first LSB, and the second bit has the value 2 and it is called the second LSB, and the third bit has the value 4 and it is called the third LSB, and the fourth bit has the value 8 and it is called the fourth LSB and so on. We have embedded the watermark text in the first LSB and also in the second LSB and in the third LSB and in the forth LSB and combined the first with the second LSB and the first with the third LSB and the first with the forth LSB and also we combined the second with the third LSB and also the second with the fourth LSB and we also combined the third with the fourth LSB. All of them will be explained in this paper. Table 1 shows the LSB uses and its maximum capacity and the size of the watermarked LSB.

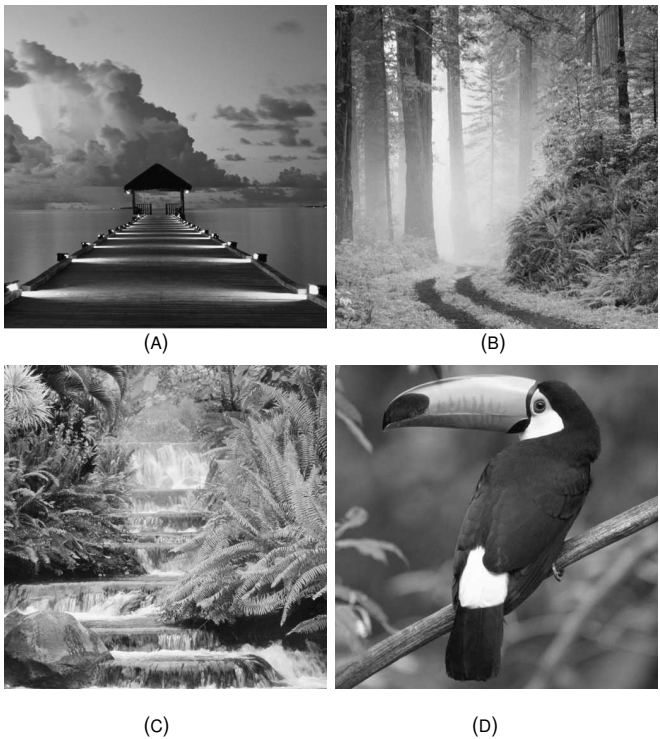

Fig. 4. The cover images: (A) Dock (B) Forest (C) Waterfall (D) Toco Toucan

### 5.1 The First LSB

Once, we embed maximum capacity of the watermark text which contains from 32766 bytes and 2 bytes to embed the length of the watermark text in determined pixels in the first LSB in the proposed algorithm and the traditional LSB [19] and Lee's algorithm [14]. Then, we got the watermarked images without noticeable distortion. The second time, we embed different watermark text which contains also from 32766 bytes and 2 bytes to embed the length of the watermark text in the four images by the proposed algorithm and the traditional LSB [19] and Lee [14] and we also got watermarked images without noticeable distortion on them. By the way, the changes in the first LSB can't be detectable by the naked eyes because the maximum change in every pixel is 1.

### 5.2 The Second LSB

In this algorithm, we embedded the same watermark text which (we embed in the first LSB) contains from 32766 bytes and 2 bytes to embed the length of the watermark text in determined pixels in the second LSB. Then, we got the watermarked images without noticeable distortion because the maximum change in every pixel is 2.

### 5.3 The Third LSB

In this algorithm, we embedded the same watermark text which contains from 32766 bytes and 2 bytes to embed the length of the watermark text in determined pixels in the third LSB and then, we got the watermarked images with some noticeable distortions in watermarked dock and Toco Toucan because the maximum change in every pixel is 4 and it is somehow noticeable.

### 5.4 The Fourth LSB

When we embedded the same watermark text which contains from 32766 bytes and 2 bytes to embed the length of the watermark text in determined pixels in the fourth LSB and, we got the watermarked images with some distortion in all watermarked because the maximum change in every pixel is 8 and the 8 grade difference is noticeable.

### 5.5 Combination First and Second LSB

When we embedded the maximum capacity of the watermark text which contains from 65532 bytes and 2 bytes to embed the length of the watermark text in determined pixels in the first and second LSB which is the proposed algorithm when the watermark text is more than 32766 bytes in this size of images, we got the watermarked images without any distortion in the watermarked images because the maximum change in every pixel is 3.

### 5.6 Combination First and Third LSB

When we embedded the same watermark text which contains from 65532 bytes and 2 bytes to embed the length of the watermark text in determined pixels in the first and third LSB and then, we got the watermarked images with some distortion in watermarked dock and Toco Toucan because the maximum change in every pixel is 5 and it is somehow noticeable.

### 5.7 Combination First and Fourth LSB

When we embedded the same watermark text which contains from 65532 bytes and 2 bytes to embed the length of the watermark text in determined pixels in the first and fourth LSB, we got the watermarked images with some distortion in the watermarked images because the maximum change in every pixel is 9 and it is noticeable.

### 5.8 Combination Second and Third LSB

When we embedded the same watermark text which contains from 65532 bytes and 2 bytes to embed the length of the watermark text in determined pixels in the Second and third LSB, we got the watermarked images with some distortion in watermarked dock and Toco Toucan because the maximum change in every pixel is 6 and it is noticeable.

### 5.9 Combination Second and Fourth LSB

When we embedded the same watermark text which contains from 65532 bytes and 2 bytes to embed the length of the watermark text in determined pixels in the Second and fourth LSB, we got the watermarked images with some distortion in watermarked images because the maximum change in every pixel is 10 and it is noticeable.

Table 1

The different uses of LSB and its maximum capacity and the size of the 512x512 BMP watermarked images

| Image | Which LSB | Watermark embedded | Size of watermarked image |
|---|---|---|---|
| dock | first | 32766 bytes | 257 KB |
| forest | first | 32766 bytes | 257 KB |
| waterfall | first | 32766 bytes | 257 KB |
| Toco Toucan | first | 32766 bytes | 257 KB |
| dock | second | 32766 bytes | 257 KB |
| forest | second | 32766 bytes | 257 KB |
| waterfall | second | 32766 bytes | 257 KB |
| Toco Toucan | second | 32766 bytes | 257 KB |
| dock | Third | 32766 bytes | 257 KB |
| forest | Third | 32766 bytes | 257 KB |
| waterfall | Third | 32766 bytes | 257 KB |
| Toco Toucan | Third | 32766 bytes | 257 KB |
| dock | Fourth | 32766 bytes | 257 KB |
| forest | Fourth | 32766 bytes | 257 KB |
| waterfall | Fourth | 32766 bytes | 257 KB |
| Toco Toucan | Fourth | 32766 bytes | 257 KB |
| dock | First and second | 65532 bytes | 257 KB |
| forest | First and second | 65532 bytes | 257 KB |
| waterfall | First and second | 65532 bytes | 257 KB |
| Toco Toucan | First and second | 65532 bytes | 257 KB |
| dock | First and Third | 65532 bytes | 257 KB |
| forest | First and Third | 65532 bytes | 257 KB |
| waterfall | First and Third | 65532 bytes | 257 KB |
| Toco Toucan | First and Third | 65532 bytes | 257 KB |
| dock | First and Fourth | 65532 bytes | 257 KB |
| forest | First and Fourth | 65532 bytes | 257 KB |
| waterfall | First and Fourth | 65532 bytes | 257 KB |
| Toco Toucan | First and Fourth | 65532 bytes | 257 KB |
| dock | Second and third | 65532 bytes | 257 KB |
| forest | Second and third | 65532 bytes | 257 KB |
| waterfall | Second and third | 65532 bytes | 257 KB |
| Toco Toucan | Second and third | 65532 bytes | 257 KB |
| dock | Second and Fourth | 65532 bytes | 257 KB |
| forest | Second and Fourth | 65532 bytes | 257 KB |
| waterfall | Second and Fourth | 65532 bytes | 257 KB |
| Toco Toucan | Second and Fourth | 65532 bytes | 257 KB |
| dock | Third and Fourth | 65532 bytes | 257 KB |
| forest | Third and Fourth | 65532 bytes | 257 KB |
| waterfall | Third and Fourth | 65532 bytes | 257 KB |
| Toco Toucan | Third and Fourth | 65532 bytes | 257 KB |

### 5.10 Combination Third and Fourth LSB

When we embedded the same watermark text which contains from 65532 bytes and 2 bytes to embed the length of the watermark text in determined pixels in the third and fourth LSB, we got the watermarked images with some distortion in all watermarked images because the maximum change in every pixel is 12 and it is noticeable.

Table 1 shows different combination of LSB for embedding the watermark. The embedded watermark text was increased when we combine 2 LSB.

## 6 PEAK SIGNAL TO NOISE RATIO (PSNR)

Notice that, there is no difference between the original and watermarked images in the first and second LSB by using our naked eyes. No distortion occurs for these watermarked images. We found some distortion when we embed the watermark text in the third and fourth LSB and also when we combined them. We got the result after we calculated the Peak signal-to-noise ratio (PSNR).

The PSNR value was used to evaluate the quality of the watermarked images. The phrase peak signal-to-noise ratio (PSNR) is most commonly used as a measure of quality of reconstruction in image compression [14]. It is the most easily defined via the Mean Squared Error (MSE) which for two mXn images I and K where one of the images is considered as a noisy approximation of the other (in other words, one is the original and the other is the watermarked image). MSE is defined as the following equation (2) and the PSNR is defined in equation (1).

$$PSNR = 10 * \log 10(\frac{MAX_I^2}{MSE}) \qquad (1)$$

$$= 20 * \log 10(\frac{MAX_I}{\sqrt{MSE}})$$

Where MAX is equal to 255 in grayscale images, MSE is the mean square error, which is defined as:

$$MSE = \frac{1}{m*n}\sum_{i=0}^{m-1}\sum_{j=0}^{n-1}[I(i,j) - K(i,j)]^2 \qquad (2)$$

Where I is the original image and K is the watermarked image.

Typical values for the PSNR are between 30dB and 40dB [14]. If the PSNR of the watermarked image is more than 30, it is hard to be aware of the differences with the cover image by the human eyes system. The cover images are shown in Figure 3. As it is explained, the invisibility of the watermark in the proposed algorithm is good. And the original image and the watermarked image cannot be distinguished by human visibility system (HVS) in some of the watermarked images. We have calculated the PSNR of all watermarked images and the result is shown in table 2 and table 3 and we have done a comparison between the proposed algorithm and the traditional LSB [19] and Lee's algorithm [14] when we embedded the same watermark text and it is shown in Table 4. The result of PSNR of the four images are more than 54 when we embed 32766 bytes as a watermark text in the second time by embedding different watermark text and we compare between our propose algorithm and the traditional LSB [19] and Lee's [14] Algorithm and we got the best results of them.

Table 2

Comparison of the PSNR of the watermarked images in the first and second and third and forth LSB

| Images | First LSB | Second LSB | Third LSB | Forth LSB |
|---|---|---|---|---|
| Dock | 54.5961 | 48.6361 | 42.5866 | 36.5826 |
| Forest | 54.6673 | 48.6080 | 42.5669 | 36.5054 |
| Waterfall | 54.6216 | 48.5651 | 42.5747 | 36.5716 |
| Toco Toucan | 54.6899 | 48.5925 | 42.5863 | 36.5180 |

Table 3

Comparison of the PSNR of the watermarked images in the combined LSB

| Images | Dock | Forest | Waterfall | Toco Toucan |
|---|---|---|---|---|
| First & Second LSB | 48.0080 | 50.8632 | 47.9522 | 47.9606 |
| First & Third LSB | 43.2265 | 43.2154 | 43.2277 | 43.2921 |
| First & Fourth LSB | 37.5494 | 37.5498 | 37.6220 | 37.6079 |
| Second & Third LSB | 41.9395 | 41.9030 | 41.9130 | 41.6953 |
| Second & Fourth LSB | 37.2002 | 37.1592 | 37.2387 | 34.2102 |
| Third & Fourth LSB | 35.8711 | 35.8387 | 35.9511 | 35.8911 |

Table 4

Comparison of the PSNR of the watermarked images Between the proposed algorithm and Lee's algorithm [14] and the traditional LSB [19]

| Image | The first watermark text which contain from 32766 bytes | | | The second watermark text which contain from 32766 bytes | | |
|---|---|---|---|---|---|---|
| | Proposed Algorithm | Lee's Algorithm | LSB Algorithm | Proposed Algorithm | Lee's Algorithm | LSB Algorithm |
| Dock | 54.5961 | 53.7041 | 53.6950 | 54.4636 | 53.8333 | 53.8282 |
| Forest | 54.6673 | 53.7650 | 53.7511 | 54.5066 | 53.8906 | 53.8720 |
| Waterfall | 54.6216 | 53.7310 | 53.7216 | 54.4895 | 53.8452 | 53.8330 |
| Toco Toucan | 54.6899 | 53.7707 | 53.7727 | 54.5402 | 53.9229 | 53.9034 |

# 7 ATTACKS ON THE WATERMARKED IMAGE

We have tested three types of attacks which are cropping and adding noise and JPEG compression in the watermarked images. The purpose of these attacks is to proof the robustness of our algorithm.

## 7.1 Cropping

We tested the watermarked images by cropping or resizing the watermarked images from 512x512 pixels to 448x448 pixels as they are shown in Figure 5. In fact, we lost some of the information after we cropped the watermarked images but we still have the information which is in the cropped images. Since the algorithm embeds many copies of the watermark text if it is not much, so we still have the information in the cropped images.

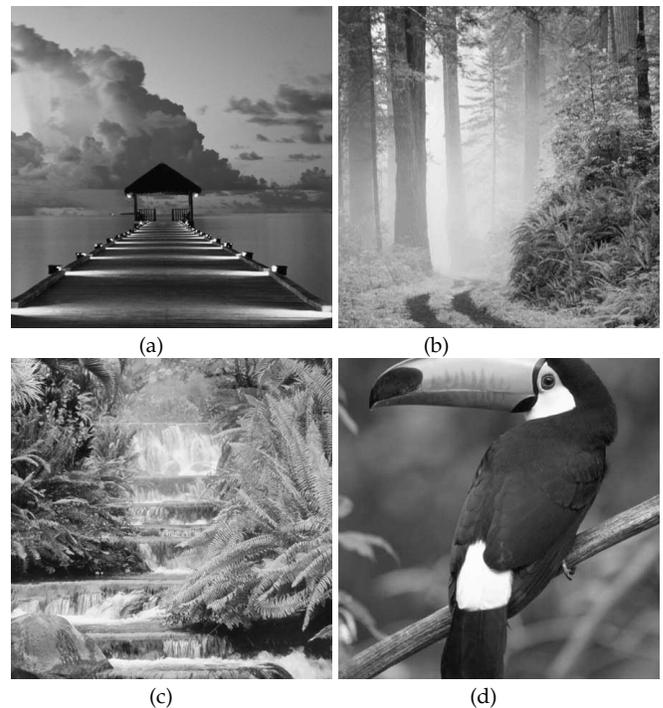

(a)                    (b)

(c)                    (d)

Fig. 5. The cropped watermarked images: (A) Dock (B) Forest (C) Waterfall (D) Toco Toucan

## 7.2 Adding Noise

We tested the watermarked images by adding noise 'salt and pepper' in the watermarked images as they are shown in Figure 6. In fact, we lost little of the watermark text which does not affect on the watermark text that much. And also, if the watermark text is not much, that will give us many copies of the watermark text. So, we can see the copies to compare the changes.

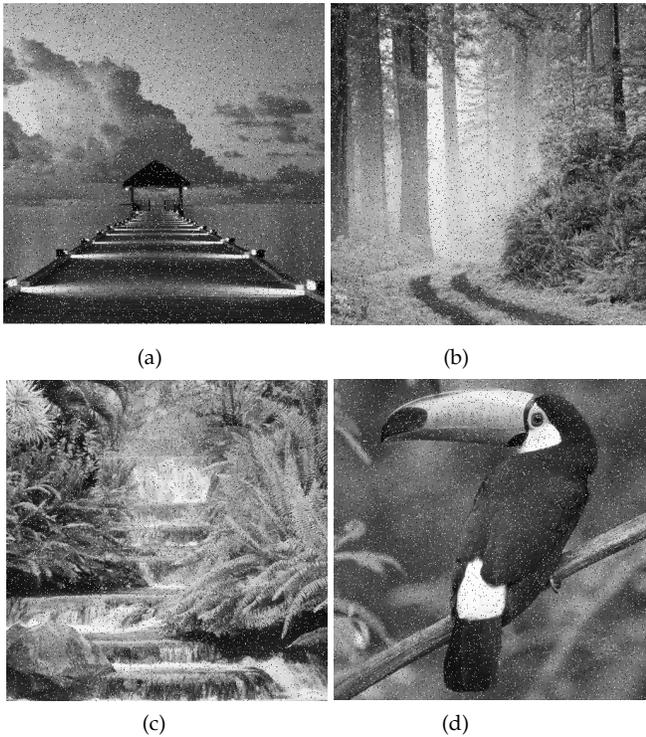

(a)             (b)

(c)             (d)

Fig. 6. The watermarked images with salt and pepper noise: (A) Dock (B) Forest (C) Waterfall (D) Toco Toucan

### 7.3 JPEG Compression

As it is known about LSB, Least Significant Bit is weak in JPEG compression because the image lost the most of LSB of the watermarked image. So, we have a simple solution of this problem which is to convert the watermarked BMP image to JPEG image. Then, we will calculate the difference between the two watermarked images by using equation (1).

$$\text{Difference} = \text{watermarked BMP image} - \text{watermarked JPEG image.} \qquad (1)$$

Then, the Difference array will be sent with the watermarked image. If the watermarked image is JPEG, we will implement equation (2) to get the watermarked in BMP format:

$$\text{Watermarked BMP image} = \text{Watermarked JPEG image} + \text{Difference.} \qquad (2)$$

After that, we can retrieve the watermark text back by using the proposed extracting algorithm from the watermarked BMP image.

## 8 CONCLUSION

This paper proposed a new LSB based digital watermarking scheme with the combination of LSB and inverse bit. The experimental result shows that the proposed algorithm maintains the quality of the watermarked image. This paper also shows the experimental results when combining different positions of LSB such as the second LSB and the third LSB and fourth LSB and the combination between them. The proposed algorithm is also tested using Peak signal-to-noise ratio (PSNR) and the result of PSNR is compared with the traditional LSB [19] and Lee's algorithm [14]. We also attack the watermarked image by using cropping and adding noise and we got good results as well. Therefore, this new digital watermarking algorithm can be used to embed watermark inside the image.

## ACKNOWLEDGMENT

This research is supported by Fundamental Research Grant Scheme (FRGS) Vote 0738.

**Abdullah Bamatraf** recived his B.Sc degree in Computer science from Hadramout University of Science and Technology, Mukalla. He is currently pursuing his study for Master degree in a New Digital Watermarking Algorithm Using Combination of Least Significant Bit (LSB) and Inverse Bit at Universiti Tun Hussein Onn Malaysia. His research area includes Image Processing and Least Significant Bit.

**Rosziati Ibrahim** is with the Software Engineering Department, Faculty of Computer Science and Information Technology, Universiti Tun Hussein Onn Malaysia (UTHM). She obtained her PhD in Software Specification from the Queensland University of Technology (QUT), Brisbane and her MSc and BSc (Hons) in Computer Science and Mathematics from the University of Adelaide, Australia. Her research area is in Software Engineering that covers Software Specification, Software Testing, Operational Semantics, Formal Methods, Data Mining, Image Processing and Object-Oriented Technology.

**Mohd.Najib Mohd.Salleh** is a senior lecturer at Faculty of Computer Science and Information Technology, Universiti Tun Hussein Onn Malaysia since 2001. He had Bachelors degree in Computer Science from Universiti Pertanian Malaysia, Selangor in 1988. He received a Master degree in Computer Science in Information System from Universiti Teknologi Malaysia in 2000. He completed his PhD in Data Mining from Universite De La Rochelle, France in 2008. His doctoral thesis was on decision tree modeling with incomplete information in classification task problem. His research interests include uncertainty in decision science, decision theory, artificial intelligence in data mining and knowledge discovery.